\documentclass[useAMS,usenatbib]{mn2e}

\usepackage{graphicx, times, ulem}
 \usepackage{graphicx}
 \usepackage{subfigure}
 \usepackage{natbib}

\usepackage{color,hyperref,topcapt}

\usepackage{amssymb,amsmath}
\usepackage{bm}
\def\red#1 {\textcolor{red}{#1}\ }   
\def\blue#1 {\textcolor{blue}{#1}\ }   

\newcommand{\apj}{ApJ}
\newcommand{\apjl}{ApJ}
\newcommand{\apjs}{ApJS}
\newcommand{\aap}{A \& A}
\newcommand{\araa}{ARA\&A}
\newcommand{\aj}{AJ}
\newcommand{\mnras}{MNRAS}

\newcommand{\nat}{Nature}

\newcommand{\Rhill}{R_{\rm Hill}}
\newcommand{\Rhillm}{R_{\rm Hill,m}}

\DeclareMathSymbol{\varOmega}{\mathord}{letters}{"0A}
\DeclareMathSymbol{\varSigma}{\mathord}{letters}{"06}

\DeclareMathSymbol{\varPsi}{\mathord}{letters}{"09}

\newcommand{\Fig}[1]{Figure~\ref{#1}}
\newcommand{\Figs}[2]{Figures~\ref{#1} and \ref{#2}}

\newcommand{\icarus}{\rm{Icarus}}

\newcommand{\ltsim}{\protect\raisebox{-0.5ex}{$\:\stackrel{\textstyle <}{\sim}\:$}}

\newcommand{\Msun}{M_{\odot}}

\newcommand{\Mjup}{M_{\rm J}}

\newcommand{\Mearth}{M_{\oplus}}

\newcommand{\lsim}{\mathrel{\rlap{\lower4pt\hbox{\hskip1pt$\sim$}}
   \raise1pt\hbox{$<$}}}                
\newcommand{\gsim}{\mathrel{\rlap{\lower4pt\hbox{\hskip1pt$\sim$}}
   \raise1pt\hbox{$>$}}}                

\title{Planet Packing in Circumbinary Systems}

\author[Kratter \& Shannon]{Kaitlin M. Kratter$^{1,2}$, Andrew  Shannon$^{3}$ \\
$^{1}$ JILA / NIST, CU Boulder, 440 UCB, Boulder, CO, 80309, USA\\
$^{2}$ Hubble Fellow \\
$^{3}$ Institute of Astronomy, University of Cambridge, Madingley Road, Cambridge, CB3 0HA, UK}

\begin{document}
\vspace{-0.5in}
\pagerange{\pageref{firstpage}--\pageref{lastpage}} \pubyear{2012}

\maketitle


\begin{abstract}
The recent discovery of planets orbiting main sequence binaries will provide crucial constraints for theories of binary and planet formation.  The formation pathway for these planets is complicated by uncertainties in the formation mechanism of the host stars. In this paper, we compare the dynamical states of single and binary star planetary systems. Specifically, we pose two questions: (1) What does it mean for a circumbinary system to be dynamically packed? (2) How many systems are required to differentiate between a population of packed or sparse planets? We determine when circumbinary systems become dynamically unstable as a function of the separation between the host-stars and the inner planet, and the first and second planets. We show that these represent unique stability constraints compared to single-star systems. We find that although the existing Kepler data is insufficient to distinguish between a population of packed or sparse circumbinary systems, a more thorough study of circumbinary TTVs combined with an order of magnitude increase in the number of systems may prove conclusive. Future space missions such as TESS provide the best opportunity for increasing the sample size.
\end{abstract}


\section{Introduction }
The {\it {Kepler}} space satellite has produced a vast new data set of dynamically interesting exoplanetary systems. In this paper, we are concerned with one of the more complex dynamical configurations observed: multi-planet, circumbinary systems.  Currently only one multiplanet circumbinary system has been discovered, Kepler 47  \citep{Orosz:2012sp}. The Kepler 47 system consists of two stars with a 3:1 mass ratio, in a $\sim 7$ day orbit, and two planets with periods of $\sim 50$ and $\sim 300$ days. It seem increasingly likely that the so-called orphan transit identified in \cite{Orosz:2012sp} is indeed a tertiary planet with a period of roughly 186 days \citep{Orosz:2013}. There are an additional 5 announced single planet circumbinary systems \citep{Doyle:2011,Welsh:2012,Orosz:2012dp,Schwamb:2013}. Although the high precision era of the Kepler mission has ended, other circumbinary planets may still be lurking in the data (\citealt{Welsh:2013}, W. Welsh, private communication).

Of particular interest is the use of circumbinary systems as a testbed for theories of planet formation.  Studies of extrasolar planets orbiting single stars show evidence that many of these systems are dynamically packed--that the planets are sufficiently closely spaced that no additional planets would be stable between them \citep{2007ApJ...665L..67B,2012MNRAS.427..770M,Fang:2013}.  Moreover, the eccentricity distribution of planets from radial velocity survey is consistent with expectations for systems  born overpacked, and reduced to their current configurations through scattering events \citep{2008ApJ...686..603J}.  Studies of the solar system also show that it is dynamically packed \citep{1996CeMDA..64..115L}.  Some have even suggested that additional planets were lost due to overpacking (e.g., \cite{2001M&PS...36..381C, 2007Icar..189..386C, 2007ApJ...661..602F, 2009Icar..204..330Y}). These lines of evidence point to a scenario where planet formation occurs around single stars with such efficacy that the systems are saturated with planets. This is the ``Packed Planetary Systems'' hypothesis.  This saturation must be explained by any planet formation theory.

Relative to the single star case, forming circumbinary planets presents additional challenges.  Processes such as secular perturbations from the binary, gas drag from an eccentric gas disk, and stochastic gravitational perturbations from the protoplanetary disk can all raise the relative velocities between planetesimals, preventing planetesimal accretion in the inner region of the disk \citep{2004ApJ...609.1065M,2012ApJ...761L...7M,2013A&A...553A..71M}.  
Although the exact extent of the region in which planet formation is suppressed remains an open question \citep{Rafikov:2013}, studies show the region to be a few to several AU in size, covering the locations of the known circumbinary planets.  Alternatively, if planet formation occurs predominantly in deadzones, circumbinary disks may have an advantage over circumstellar disks due to the reduction in accretion by the action of the binary torque on the inner disk edge \citep{Martin:2013}.  Depending on the ease with which planets are formed, and the range of allowed radii,  these hurdles may be significant enough that circumbinary planetary systems are dynamically sparse, rather than dynamically packed.  Either outcome would represent a significant observational constraint on models of planet formation.

To understand how planet formation proceeds around binaries, we begin by comparing the population of planets around binaries to that of single stars. Computing the frequency of circumbinary planet occurrence is insufficient as we do not have evidence that circumbinary protoplanetary disks are nearly so universal as circumstellar ones. Recent work shows that the overall  disk fraction for binaries is lower \citep{Harris:2012}, though this work does not address the properties of circumbinary disks specifically.  Additionally, the mechanism by which the host stellar systems form remains uncertain \citep{Tohline_binrev}. One of the most commonly invoked formation mechanisms for close binaries is ruled out in these cases: Kozai circularization via Tidal Friction (KCTF) \citep{Fabrycky:2007}. KCTF requires that a distant third body drive the binary to high eccentricity, where the stars interact tidally, leading to orbit circularization on Gyr timescales. Not only does the binary achieve a close orbit on timescales long compared to protostellar disk lifetimes, it also would have been on an orbit plunging through the planet forming region on the way to its current location. Thus the existence of these systems places strong constraints on the processes by which the host stars form themselves. 

In the absence of information on the natal disks, an alternate approach to understanding the efficacy with which circumbinary disks make planets is to examine their packing properties in comparison to those of single star systems.  By considering only systems with at least one planet, we are assured of only considering systems which did possess protoplanetary disks, and systems which have not lost their planets to binary evolution processes such as KCTF.  In this work we entertain the ansatz that the ease with which disk material is converted into planets scales with the packing of the system.

The only other reference point for circumbinary multi-planet systems resides in our own solar system. The Pluto-Charon planet-satellite system consists of a binary orbited by 4 low mass satellites. The dynamical stability of this system was recently studied by \cite{YKK2012}. They found that circumbinary multi-planet orbital stability cannot be approximated by the constraints on either single-star multi-planet systems, or single-planet circumbinary systems.  

In this paper we first constrain circumbinary, multi-planet stability by exploring the minimum stable intra-planet separation as a function of star-inner planet separation. These two limits, investigated by \cite{HW99} and \cite{Gladman:1993} respectively, are interdependent in the circumbinary, multi-planet case. Close inner planets can only remain stable with more widely spaced secondaries.  In Section \ref{sec:def}, we conduct a numerical parameter study of two-planet circumbinary systems. In Section \ref{sec:kep47} we examine the specific case of Kepler 47, and ask whether the system is dynamically packed with the three (likely) planets. Finally, in Section \ref{sec:monte-carlo}, we make a simple estimate for the detection frequency of secondary (or tertiary) planets in circumbinary systems, in order to estimate how many detections are required to reliably distinguish between a packed and sparse population. We discuss the implications of our findings, and make suggestions for future observational endeavors in Section \ref{sec:discussion}.

\section{Defining a Packed Circumbinary System} \label{sec:def}
To compare with single-star systems, we must first determine what constitutes a packed circumbinary system.

\subsection{Previous Constraints}
We extend the work of three well known dynamical studies: \cite{HW99}, \cite{Gladman:1993}, and \cite{Chambers:1996}. \cite{HW99} systematically explored the minimum semi-major axis for stable orbits around either a single star in a binary (s-type orbits) or about both components of a binary (p-type orbits). They found that the critical semi-major axis ratio between the planet and stars varied with stellar mass ratio and eccentricity. For p-type orbits, with which we are concerned in this work, they found a minimum stable semi-major axis around equal mass, circular binaries of $a_{\rm crit} \equiv a_p/a_*$ = 2.3, where $a_p$ is the planetary semi-major axis and $a_*$ the binary semi-major axis.

\cite{Gladman:1993} showed analytically that there is minimum separation between two planets orbiting a single star, where Hill stability is guaranteed; Hill Stability guarantees that there will be no close encounters between the two bodies. The analysis assumes a large mass ratio between the star and planets. The critical intra-planet separation can be measured in terms of the planetary Hill radius:
\begin{equation}
 \Rhill= \left(\frac{m_p}{3M_*}\right)^{1/3}a_p
\end{equation}
\cite{Gladman:1993} finds that the minimum stable separation, in units where $M_*=a_p = G = 1$, is $\Delta_c = 2.4*(\mu_1 +\mu_2)^{(1/3)}$, where $\mu_1$ and $\mu_2$ are the planetary masses. This corresponds to a minimum spacing of $\sim 3.5\Rhill$, for equal mass planets.

It is convenient to introduce the concept of a mutual Hill radius, as this scale incorporates both planetary masses in estimating the strength of planet-planet perturbations. The mutual Hill radius is defined as\footnote{The mutual Hill radius is a somewhat ill-defined metric for arbitrarily large planet separation. Because of the dependence on both the inner and outer planet semi-major axis, for a given system there is a maximum number of mutual Hill radii by which a system can be separated because the outer planet separation increases faster than the number of Hill radii in between the planets.}:
\begin{equation}
\Rhillm = \left(\frac{m_{p1}+m_{p2}}{3M_*}\right)^{1/3}\left(\frac{a_1+a_2}{2}\right)
\end{equation}

 The work of \cite{Chambers:1996} considers single-star systems with 3 or more planets.
While an analytic solution is not possible, numerical work has shown that the average system lifetime increases with increasing spacing. Stability over Gyr timescales requires spacings greater than $ \approx 8-10\Rhillm$  \citep{Chambers:1996}. In addition, the timescale on which a system spaced by a fixed number of $\Rhillm$ becomes unstable decreases with increasing number of planets up to $5$, and then plateaus \citep{Chambers:1996,Smith:2009}.
As shown by \cite{YKK2012}, none of the above limits accurately describe the stability of multi-planet circumbinary systems.

\subsection{N-body integrations}

We conduct a series of N-body integrations to constrain the minimum separation allowed between two planets in orbit about a binary, as a function of the distance of the innermost planet from the binary. We use the publicly available n-body code, {\it {Swifter}} to carry out all of our calculations \citep{Levison:2013}. We use the Gauss-Radau 15th order integrator as this is relatively efficient, while making no assumptions regarding the system architecture. See the appendix of \cite{YKK2012} for a discussion of the numerical issues associated with this integrator.

We vary two parameters: $a_{\rm in}$, the distance of the innermost planet from the stellar barycenter in units of the stellar semi-major axis, and $\beta$, the spacing between the planets in units of mutual Hill Radii. We explore $2.5 < a_{\rm in} < 3.5$ and $ 3.2 < \beta <  9$. The step size in $a$  is 0.25, and for $\beta$ as small as 0.05\footnote{We have chosen to selectively plot values of $\beta$ to simplify the figures. For the lowest mass case, we increased the $\beta$ spacing once monotonic trends became apparent.}. For each $a_{\rm in}$ and $\beta$, we run 100 different realizations of the planetary system with random phases relative to each other and to the binary. For simplicity we use circular, Keplerian, co-planar orbits. Note that such orbits are not precisely circular as the forced eccentricity of the binary can be $O(10^{-2})$ \citep[see e.g.,][]{Mudryk:2006,YKK2012}.  We consider equal mass stars, and a planet- primary star mass ratio of $\mu=10^{-3},10^{-4},10^{-5}$. The minimum $a_{\rm in}$ is just outside of the test particle stability limit of \cite{HW99}, while the minimum $\beta$ is just inside the minimum separation for two planets to be Hill stable about a single star. All of the observed systems have inner planets outside of our minimum $a_{p,1}/a_{\rm crit}$. We focus on the results for the two lower mass ratio cases; as discussed below, the Jupiter mass planets have nearly identical limits to the single-planet / single-star case.

Each of the systems is integrated for over $2 \times 10^8$~binary orbits. A system is deemed unstable if either of the planets crosses the stellar orbit, or is ejected from the system. We use the median lifetime of all realizations to compute the instability timescale. 

\subsection{Numerical Results}
\begin{figure}
   \centering
\includegraphics[scale=0.7]{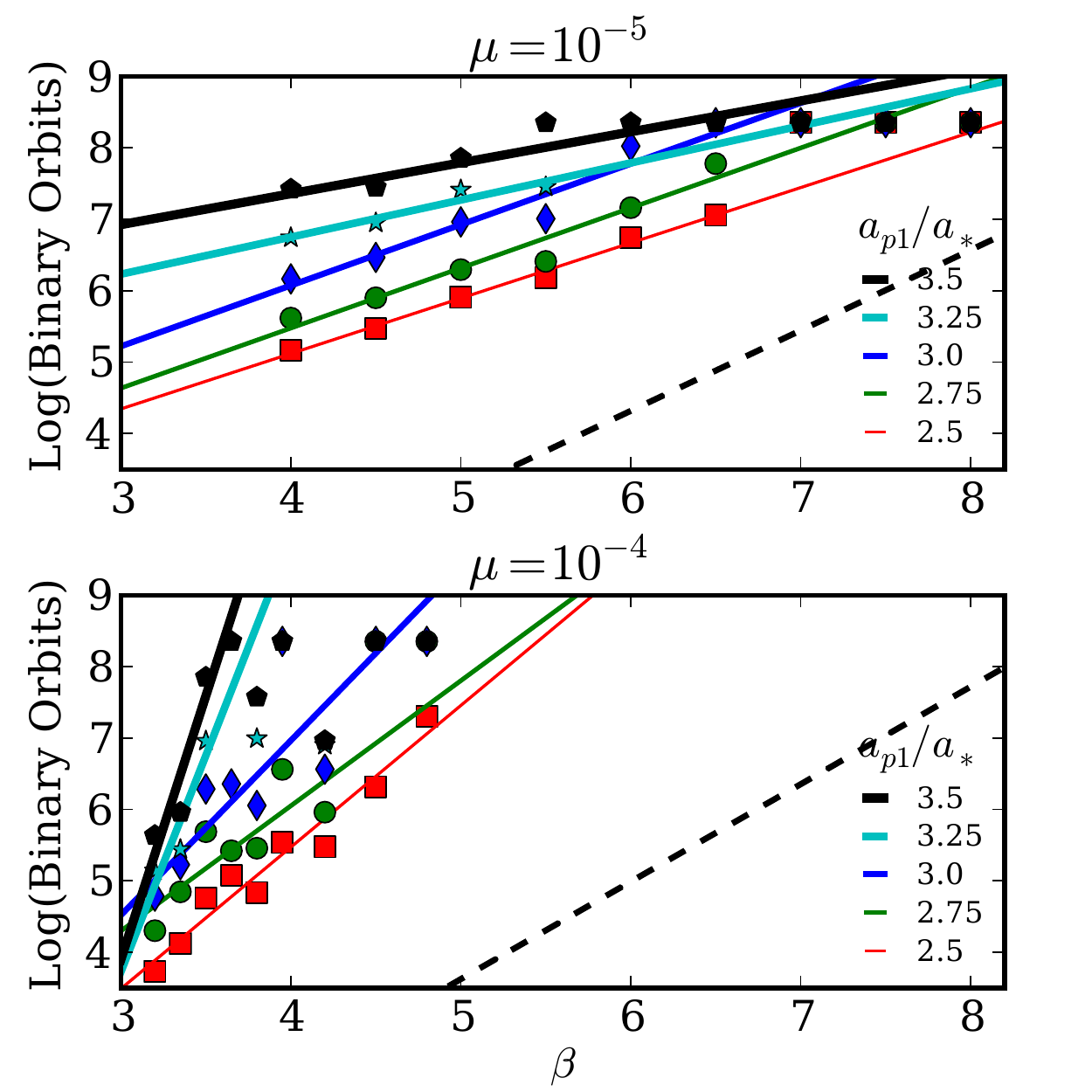} 
   \caption{Instability timescale as a function of $\beta$ for different inner planet separations.  The panels are labeled by the planet-star mass ratio.  The power law fits are quite good for the low mass case (top), whereas higher mass planets (bottom) show non-monotonic stability trends. For lower mass planets, the scaling of stability with $\beta$ is similar for all separations, whereas the more massive planets become more stable at lower $\beta$ very quickly with increasing $a_{p,1}$. The black dashed line indicates the fit for three planets from Youdin et al (2012) for the data of Chambers (1996). }
 \label{tvsbeta}
\end{figure}

As seen in \Figs{tvsbeta}{tvsa}, our results are in line with expectations qualitatively: when the inner planet separation is very small, the planets are easily destabilized when $\beta$ is small.  When $a_{\rm in}$ is large, the second planet can remain stable near the critical $\beta =3.5$ of \cite{Gladman:1993}.
The spacing required for stability increases steeply with decreasing $a_{\rm in}$, although the two planets remain somewhat {\it{more}} stable than three planets around a single star at fixed intra-planet separation (for the mass ratios considered here). 

In some ways this last trend is not surprising: the inner planet is typically $20-30 R_{\rm hill}$ from the barycenter of the stellar orbit. To the extent that the secondary star acts like a massive planet, the system is very well spaced. The most naively surprising trend is that in order to recover the two planet, single-star limit, lower mass planets must be moved further away from the binary than their higher mass counterparts; Jupiter mass planets only 20\% outside the single planet critical stability radius are stable when spaced by $3.5\Rhillm$. When the mass ratio drops to $\mu=10^{-4}$, the critical inner planet separation increases to $\sim 60\%$ of the single planet limit.  Going down another order of magnitude, Earth mass planets must be placed roughly $75\%$ further out from the binary to recover the single-star, two planet packing limit. This increase is explained by the larger physical separation of the second planet from the binary: because the Hill radius is larger for more massive planets,  the outer planet is ~20\% further out for Jupiter mass planets than for Earth mass ones. Since the quadrupole moment from the binary falls off like $r^{-3}$,  it has decreased by 50\% at the location of the second Jupiter mass planet compared to the second Earth mass planet.

In order to compare with the previous 3-body results, we fit our data to the following power law formula based on \cite{Quillen:2011} for instability generated by three-body resonances:
\begin{equation}
\rm{log}(t_{\rm in})  = k_1 + k_2\beta\mu^{1/12} + k_3\rm{log}(\mu)+k_4\frac{a_{p,1},}{\rm AU}
\end{equation}
where $t_{\rm in}$ is in years, and we have added an additional term to include the inner planet semi-major axis. Our best fit is:
\begin{equation}
\rm{log}(t_{\rm in})  =  -7.75 + 3.05\beta\mu^{1/12} -0.28\rm{log}(\mu)+1.88\frac{a_{p,1}}{\rm AU},
\end{equation}
For comparison, the fit to the Chambers results gives $ k_1=-9.11,k_2 = 4.39, k_3 = -1.07$ (where there is no $k_4$).

In \Figs{tvsbeta}{tvsa}, we show two-dimensional fits for t as a function of $\beta$ and $a_{p,1}$  only to demonstrate the applicability of power law scalings. 

 The increasing scatter with mass is consistent with the single star, 3-planet results \citep{Chambers:1996,Smith:2009}. Clearly for the typical masses of circumbinary planets, it is difficult to accurately predict the stability of a given system based on a simple logarithmic scaling.
 
A more in depth exploration of the stability of two-planet systems around a wider variety of binary configurations will be the subject of future work. The precise stability of any given system architecture may not be captured by $a_{\rm in}$, and $\beta$ alone. The same is true for single star systems. For example, Kepler 11 \citep{Lissauer:2009} has pairs of planets which would violate the three-planet stability criteria, but they are sufficiently distant from other planets in the system that they remain stable. To understand whether any given system is truly packed requires individual simulations. Nevertheless,  we can see that for the masses with which we are concerned ($\Mearth-\Mjup$), the circumbinary systems are typically at least as stable as 3-5 planet systems around single stars, so long as the inner planet is more than a factor of $\sim1.2$ outside of the single planet critical radius. One approaches the single star, 2 planet limit outside of $a_{p,1}/a_{\rm crit}= 1.5$. All of the observed single planet systems lie in the intermediate regime of $1.1 \ltsim a_{p,1}/a_{\rm crit} \ltsim 1.5$.  These assumptions allow us to predict the number of expected detections of secondary planets in packed and sparse circumbinary systems in Section \ref{sec:monte-carlo}.

 An alternative approach is to study individual systems, thus, we now turn to the specific case of Kepler 47, and subsequently the currently known single planet systems. 
 
\begin{figure}
   \centering
   \includegraphics[scale=0.7]{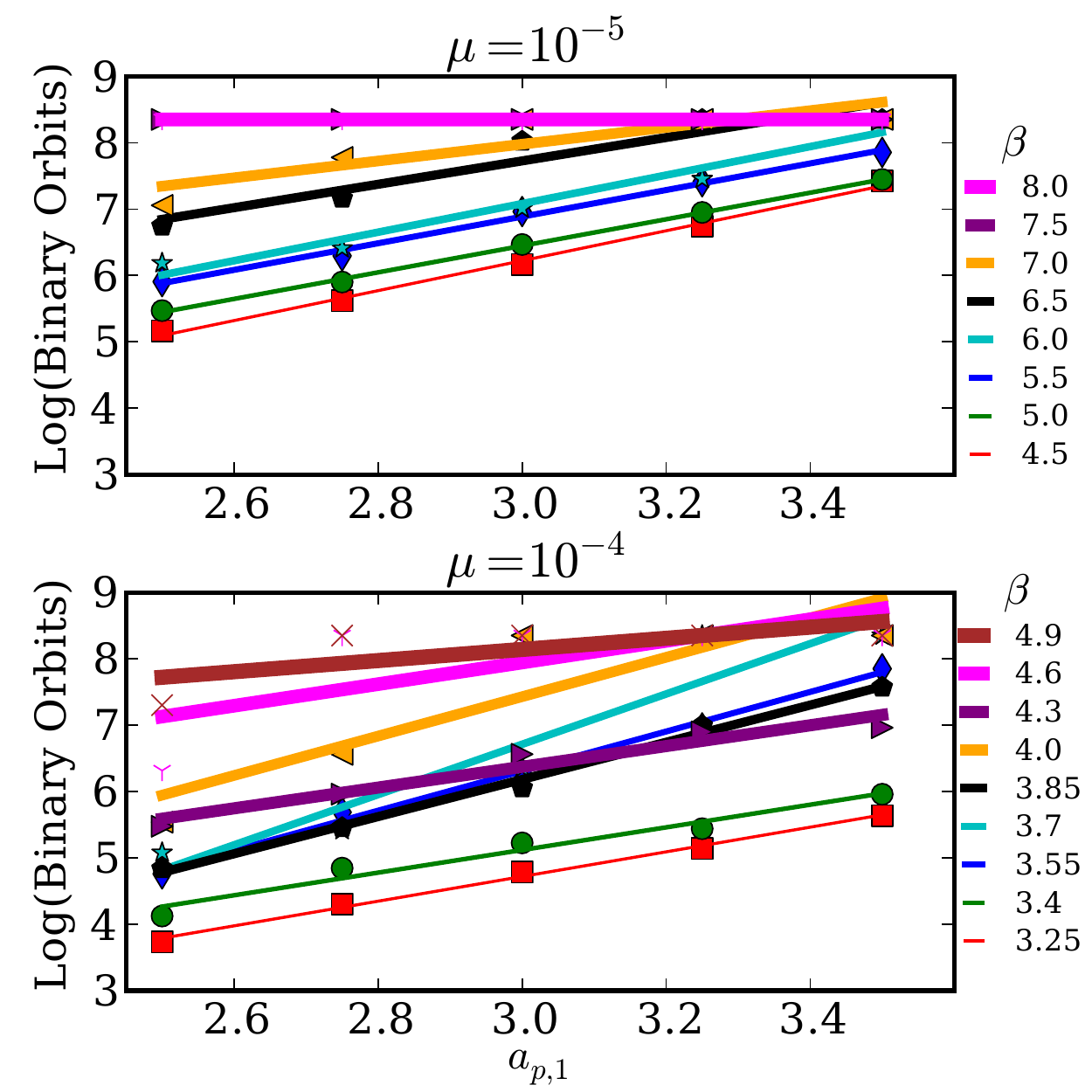}
      \caption{The same data as \Fig{tvsbeta} but now showing instability timescale as a function of inner planet semi-major axis, $a_{p,1}$ for different values of $\beta$. Note that lower mass planets are less stable than higher mass planets for the same parameters.} \label{tvsa}
\end{figure}

\section{Case Study: Kepler 47}\label{sec:kep47}
Kepler 47  is thought to contain three transiting planets, with Kepler 47d being more uncertain (Orosz et al, in prep). Kepler 47d is predicted to orbit between planet b and c, with a period of roughly 186 days. For a mass comparable to c, it is dynamically packed with respect to planet c: it has a separation of roughly $\beta = 10$. By contrast, the separation between planets b and d is rather large, about $\beta = 28$. Unlike the single planet systems, the inner planet (b) is relatively well separated from the binary with $a_{p,1}/a_* \approx 3.5$, which for a binary mass ratio of $M_2/(M_1+M_2) \approx 0.25$ translates to $a_{p,1} = 1.57~ a_{\rm crit}$ \citep{HW99}.  Based on our integrations in Section \ref{sec:def}, there is in principal place for a planet in between b and d, ignoring any contribution from c, especially considering the wide inner spacing. However, comparing the stability of two and three planets in the single star case suggests that the required distance will be much greater. Our {\it{two planet}} fit predicts that the lifetime of a planet intermediate between b and d, at $\beta=14$ is $10^{14}$ years.

 In order to gauge the likelihood of such a stable orbit, and quantify whether the system is likely to be packed, we run an integration of the Kepler 47 system to check for stable interior orbits.

\subsection{Orbital Parameters}
We use one of the best fit models from the study of \cite{Orosz:2012sp} to determine the initial orbital configuration for the binary, and planets b and c (J. Carter, private communication). Note that for the existing systems the constraints on eccentricity of the outer planet are limited. We place planet d on a 186 day orbit with zero initial eccentricity. While the fits to the Kepler light curves assumed massless planets, we assign the planets masses based on their inferred radii and the  planetary radius-mass relationship in \citep{Lissauer:2011fy}. The initial state vectors. planetary masses, and orbital elements are listed in Table  \ref{tab:statevecs}. 
\begin{table*}
   \centering
   \topcaption{ State vectors for Kepler 47 system with the primary at the origin with zero velocity. Mass is in units of $\Msun=1$, $G=1$, and length units are in AU.} 
   \begin{tabular}{cccccccccc} 

      $M/M_\odot$ &$ x$&$y$&$z$&$v_x$&$v_y$&$v_z$&$a$&$e$&$i$ \\
     \hline
     1.043 & 0.0 & 0.0 & 0.0 & 0.0 & 0.0 & 0.0 & 0.083& 0.032 & 0.0\\
               0.362   & -8.638e-2 & 1.554e-4 & 0.0 &-7.379e-3  & -3.966 &  0.0 & 0.083& 0.032 & 0.0\\
      2.985e-5 &-0.146 &   -0.271    &  1.338e-3 &   1.971 &      -1.90 &      2.892e-3 & 0.293& 0.0127& 0.004\\
       7.295e-5 & -0.649 &       0.347 &     2.843e-8 &  -0.680 & -2.248 &     6.115e-3 & 0.712 & 0.0 & 0.0\\
        7.295e-5 & -1.575e-2 & -0.985 &     1.611e-2 &  1.173  &   -0.797 &     1.057e-2 &0.947& 0.181& 0.008\\
         \hline
   \end{tabular}

   \label{tab:statevecs}
\end{table*}

To search for stable orbits we populate the region in between Kepler 47b and d with 1000 test particles on circular, coplanar orbits about the system barycenter. We set the minimum and maximum separations of the test particles to be $5R_H$ outside and inside of planets b and d respectively.  The spacing limits are chosen conservatively based on our two planet integrations above. 

\subsection{Results}
\begin{figure}
   \centering
      \includegraphics[scale=0.7]{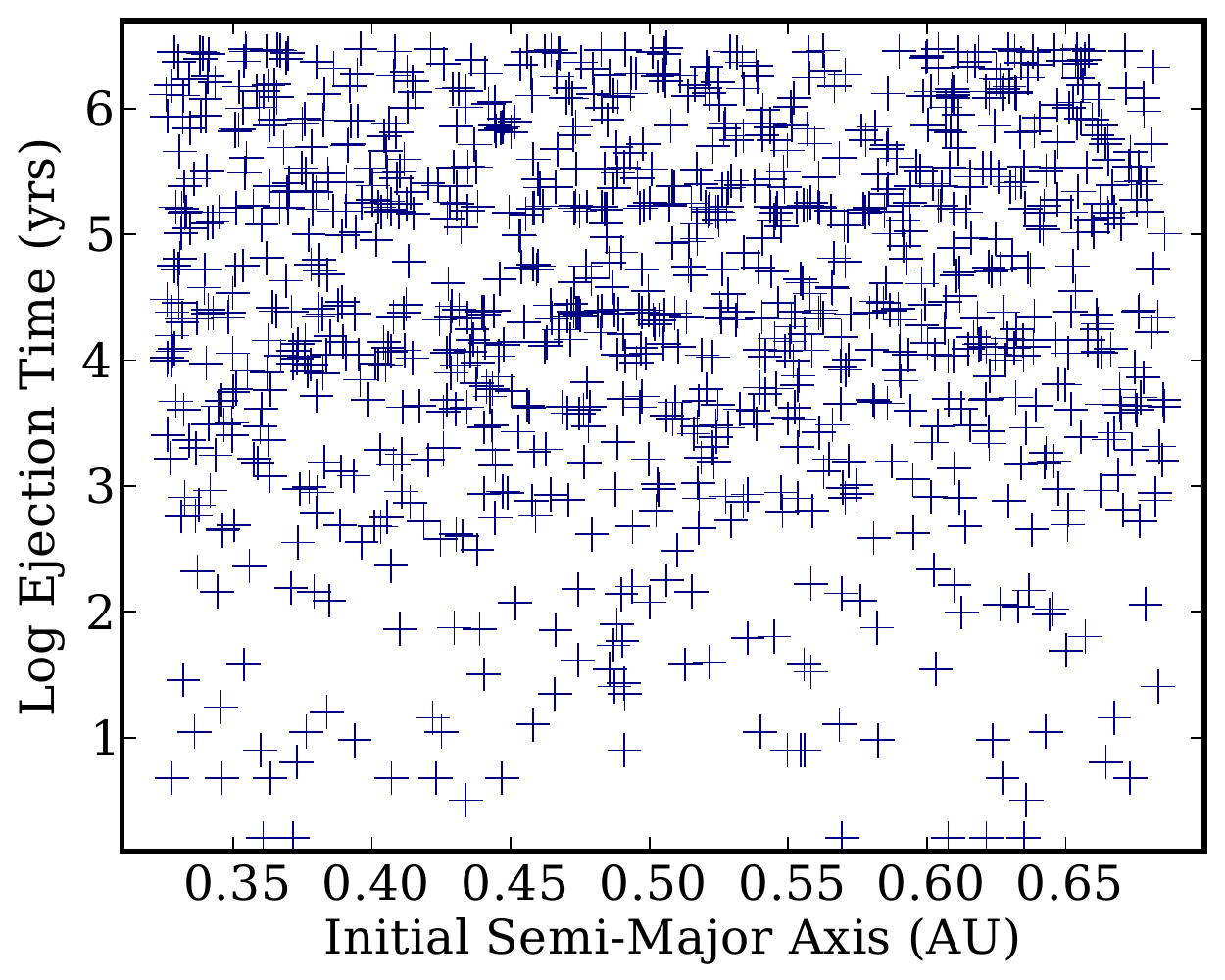}
          \caption{ The distribution of test particles ejected from orbits between Kepler 47b and orphan planet Kepler 47d, as a function of initial semi-major axis. There appears to be a slight increase in stability at early times for orbits roughly equidistant from b and d.}
   \label{fig-kep47a}
\end{figure}

After $2\times 10^6$~years (or roughly $10^8$~binary orbits) all of the test particles were lost due to crossing the orbit of either planet b or d. We show both the distribution of particle lifetime with semi-major axis and the distribution of particle ejection in time in \Figs{fig-kep47a}{fig-kep47t}. For the masses and orbital parameters chosen, planet c has both a large initial eccentricity, and also a somewhat variable orbit. Based on the variability in planet c, we also predict that either the mass of planet d or the eccentricity of planet c is lower. Although we have clearly not sampled the full phase-space of intermediate orbits, all of those chosen have lifetimes far shorter than the system age. Thus we conclude that with planet d in place, Kepler 47 is  most likely dynamically packed. To the extent that $\beta$ controls stability, changing the planetary mass by a factor of $2$ only shifts $\beta$ by $\sim 1.25$, and thus our results are likely to hold for reasonable masses of planet d.

\begin{figure}
   \centering
              \includegraphics[scale=0.7]{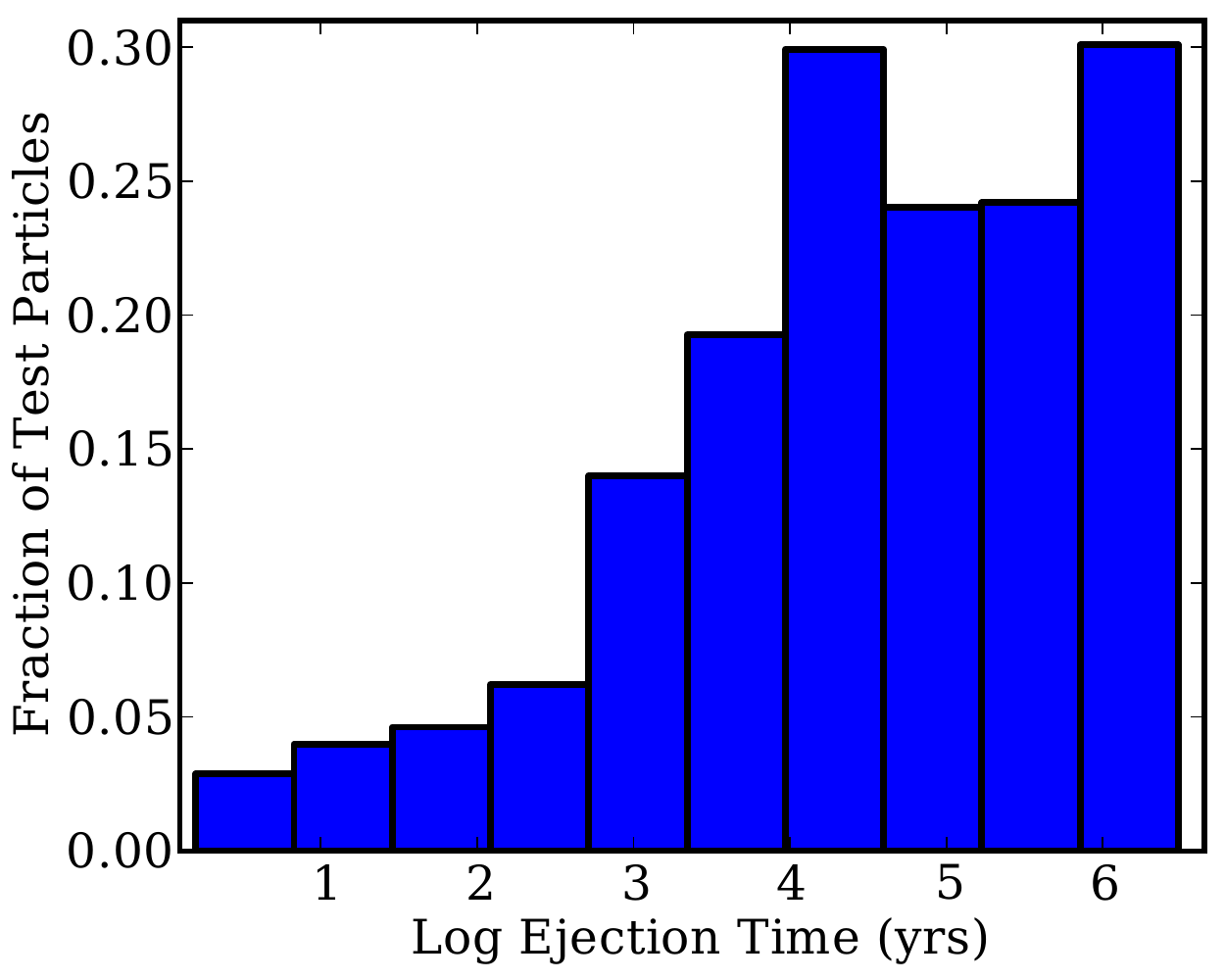}
   \caption{Distribution of particle ejections as a function of time. The majority of particles are lost by a few $10^5$ years. The rapid ejection of particles compared to the system lifetime suggests that it is packed with three planets.}
   \label{fig-kep47t}
\end{figure}

\section{Interpreting Existing Systems: Are Circumbinaries Packed?}\label{sec:monte-carlo}

Is the packing in Kepler 47 representative of the population as a whole?  Clearly our investigation is hindered by poor statistics, but using the  results from our n-body integrations we can quantify how many systems are required to distinguish between a packed and sparse population.

As demonstrated above, we can roughly consider circumbinary two-planet systems packed if they have similar spacings to packed 3-5 planet single-star systems.
The nominal {\it minimum} separation according to \cite{Chambers:1996} or \cite{Smith:2009} is of order $\beta = 8-10$.  One should therefore expect subsequent planets to be located between $\sim8 < \beta < 20$ away from each other. According to \cite{Fang:2013}, the Kepler single star multiple systems are consistent with being packed in $\sim 35-45\%$ of the known systems with 3-4 planets. The distribution of separations measured in $\beta$ is a Rayleigh distribution with $\sigma \approx 17$.

Using the Kepler single star sample as an example of a packed population, we now pose the question: how many multi-planet circumbinary systems should {\it{Kepler}} have observed in the current sample if the circumbinary systems are equally packed? If they are sparse?

\subsection{Observability of secondary planets via transits}
\begin{figure}
   \centering
   \includegraphics[scale=0.25]{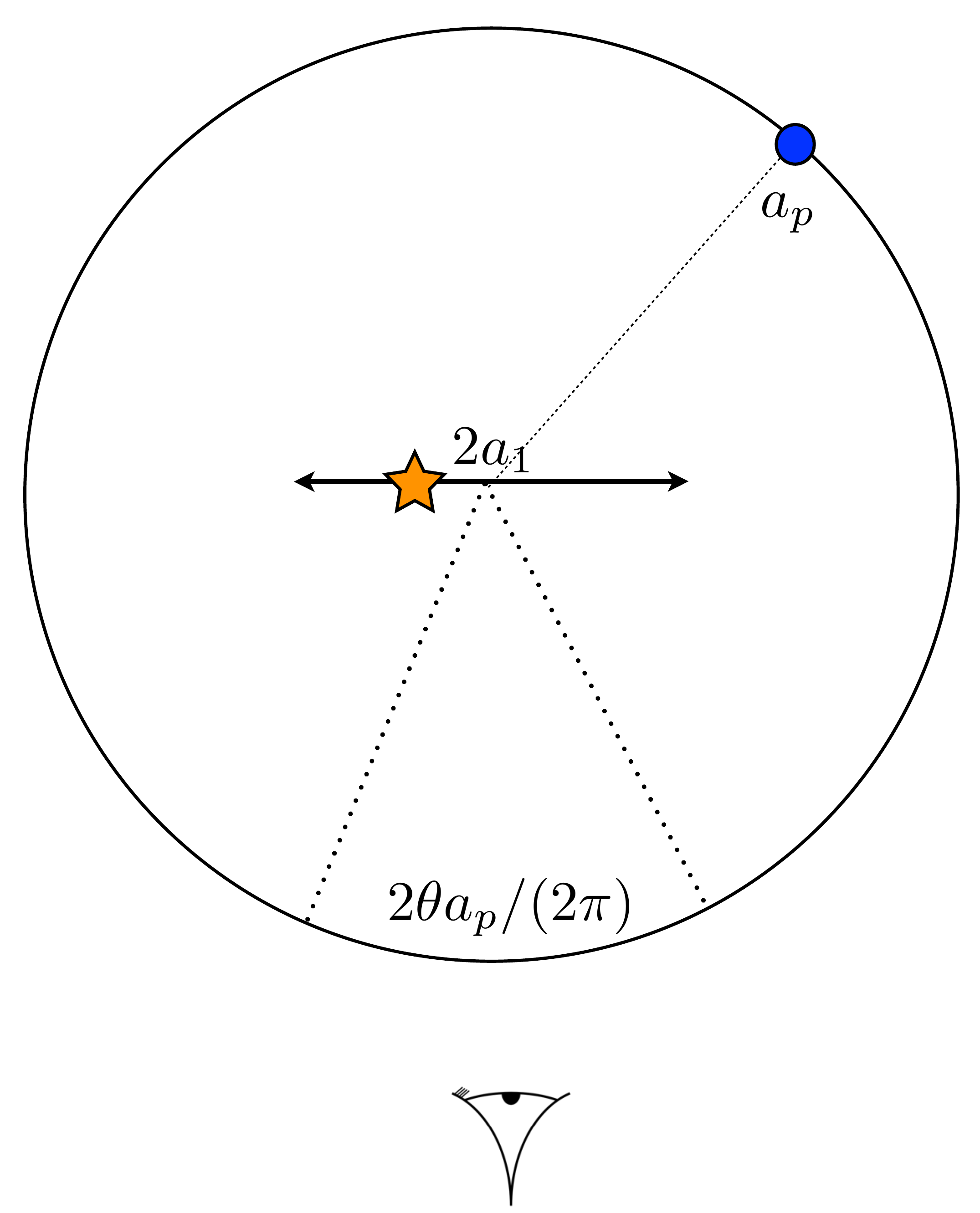}
      \includegraphics[scale=0.25]{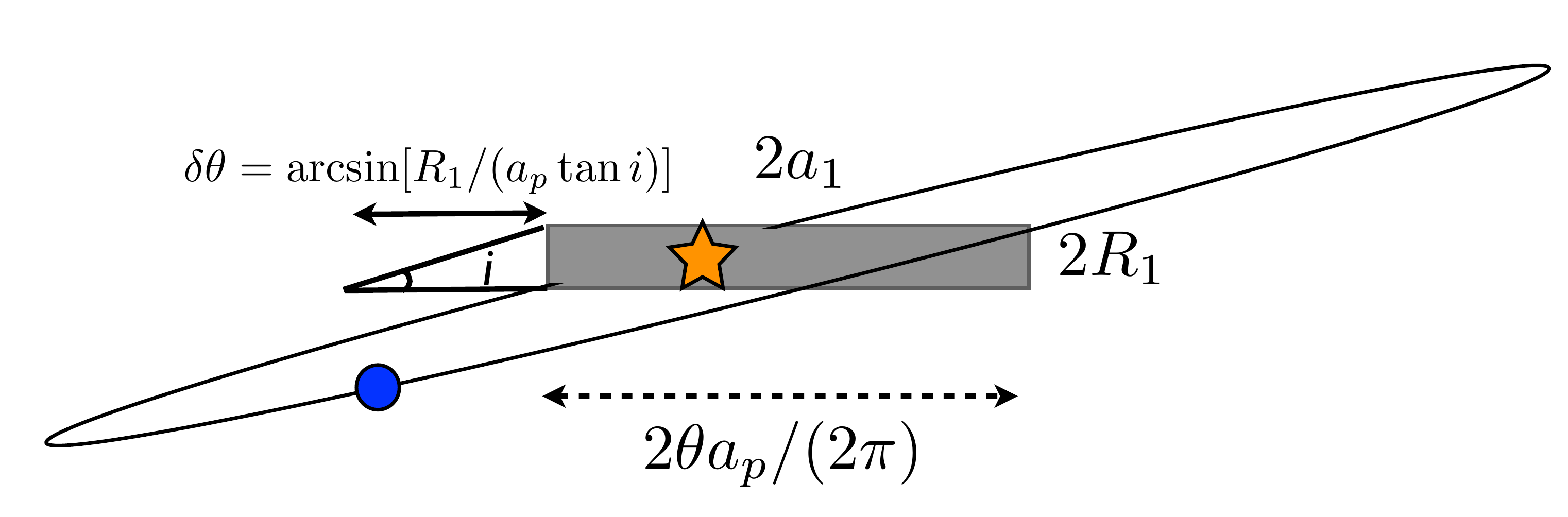} 
   \caption{Schematic diagrams. Top panel shows birds eye view of a circumbinary system with the project transit arc, $2\theta$, while the bottom shows the edge on view, illustrating the extra $\delta\theta$ added to account for some certain combinations of inclinations and lines of ascending node. }
   \label{transit_draw}
\end{figure}

Before considering an ensemble population, we need to estimate a rough transit probability $P_{\rm tran}$ for a second planet in an already detected circumbinary system. We make the simplifying assumption that the binary itself is  edge-on (consistent with the star eclipsing), and that any transits will occur over the primary star. From the point of view of the observer, consider a rectangle projected on to the plane of the sky which encompasses all possible locations of the primary star, with a height twice its radius ($R_1$). This rectangle has dimensions 
\begin{equation}
A_r = 2a_*\frac{m_2}{m_1+m_2} \times {2R_1}= 2a_1 \times 2R_1,
\end{equation}
where $a_1$ now refers to the primary's orbit about the barycenter. If the mutual inclination of the planet and binary is zero, the planet will always transit. As the inclination increases, transits will only occur if the location of the line of ascending nodes guarantees that the projected cord of the planet orbit crosses the projected rectangle of the stellar orbit (we shall multiply by 2 later for descending nodes in this arc as well). \Fig{transit_draw} illustrates the geometry we consider.
To determine the fraction of allowed ascending nodes, consider the fraction of the planetary orbit, which intersects the stellar rectangle: 
\begin{equation}
2\theta = 2\arcsin{\left(\frac{a_1}{a_p}\right)}.
\end{equation}
This angle encompasses most allowed values for the ascending node. However, for small inclinations, there will be an angular offset on either side of $\theta$ where an ascending node location will still send the planet grazing across the top or bottom of the star.  These two segments have size 
\begin{equation}
\delta\theta = \frac{R_1}{a_p\tan(i)}.
\end{equation}

Therefore, the total probability (including descending nodes) that such a planet will ever transit the star is:
\begin{equation}
P_{\rm trans} = 2\frac{2\theta +  \delta\theta}{2\pi} = \frac{1}{\pi}\left[2\arcsin{\left(\frac{a_1}{a_p}\right)}+\arcsin{\left(\frac{R_1}{a_p\tan{i}}\right)}\right]
\end{equation}

We use this simplified probability to estimate whether or not a hypothetical secondary planet should be observed to transit. Circumbinary systems with eclipsing binary stars have a much higher transit probability at a given inclination and separation. When considering secondary planet transits only, we have already specified that the stars are eclipsing, and thus even relatively high inclination planets should be expected to  cross the  area swept out by the stars. It is only in the case where one specifies that the binary eclipses that the probability greatly exceeds that for the single star case, as noted by \cite{Borucki:1984}. Note that we have {\it{only}} defined the probability that a transit is possible in infinite time. Depending on orbital inclination, the probability of transit {\it{per orbit}} can differ significantly from unity. We do not include this in our calculations at this time, although we do remove planets with periods longer than the Kepler mission lifetime from the detection statistics. In this aspect our calculations are optimistic, although we also pessimistically exclude the possibility of multiple transits per orbit, and transits of the secondary \citep[which have been observed in Kepler 16 by ][]{Doyle:2011}.

\subsection{Observability of secondary planets via TTVs}
In the absence of (or in addition to) transits, one can also infer the existence of an outer massive body based on its gravitational effect measured through variations in the occurrence time of the transits, so-called TTVs. While there are enormous TTVs  due to the stellar motion,  there may also be smaller TTVs due to an outer planet. \cite{Holman:2005} estimate the magnitude of the change in transit time occurrence in single star systems to be:
\begin{equation}
\Delta t \approx \frac{45 \pi}{16}\frac{M_{p,2}}{M_*}P_1\alpha_e^3(1-\sqrt{2}\alpha_e^{3/2})^{-2}
\end{equation}
where $\alpha_e = a_{p,1}/[a_{p,2}(1-e_2)]$. We emphasize that this does not include any perturbations due to the binary. However, the TTVs due solely to the motion of the binary as a moving target can in principal be subtracted given a precisely characterized stellar orbit. TTVs due to the binary potential itself should have a significantly higher frequency modulation than that due to an outer planet (D. Nesvorny, private communication). We thus take this as an order of magnitude approximation for the amplitude of the variation induced by a second planet. A rigorous investigation of the detectability of TTVs due to an outer perturber in circumbinaries is beyond the scope of this paper, and will be the subject of future work.

\begin{figure}
   \centering
              \includegraphics[scale=0.65]{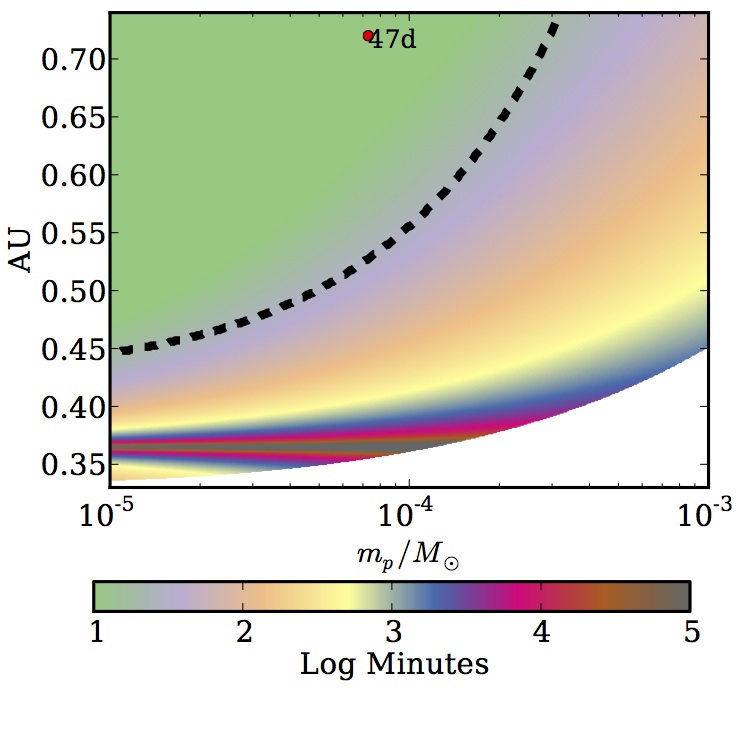}
   \caption{Magnitude of predicted TTV signal in log(minutes) for an outer planet perturbing the orbit of Kepler 47b, assuming a zero eccentricity orbit. The location of the orphan transit (47d) is shown for reference. The white region shows orbits prohibited by stability, and the black dashed line shows where $\beta=20$, delineating packed and sparse orbits based on the two planet case. Most packed systems would have TTV signals greater than 1 hour.}
\label{ttv_mag}
\end{figure}

Because the orbits are long compared to a typical transiting planet, so too are the TTVs. In \Fig{ttv_mag}, we show the magnitude of the TTV's induced on Kepler 47b as a function of the mass and semi-major axis of a perturbing outer planet (assuming $e=0$). The black dashed lines indicates the boundary in mass and semi-major axis above which the two planets  would be separated by more than $\beta=20$. In this case, the detection of a TTV above $\sim1$ hour would indicate a dynamically packed 2 planet system, even without characterizing the mass or orbit of the perturber. Thus even without solving the inverse problem, for a given system it is possible to correlate TTV magnitude with planet packing directly.

Based on the stated precision of the single planet transit times of roughly 5 minutes \citep[see, e.g., table S4 of the supplementary information of][]{Orosz:2012sp}, we adopt 20 minutes as the required size of the TTV to be detectable. For comparison, we also consider the detection probabilities for TTV signals that are at least 2.5 hours. Note that based on this formula, Kepler 47b would have TTVs of order 5 minutes due to the posited middle planet d, which have not been reported. Meanwhile, planet c would induce 18 hour TTVs on planet d, which we would expect to be both detectable and frustrating to orbit fitting with only a few transits.
For most of the current sample of CBPs, the paucity of transits is likely the limiting factor in detecting TTVs: uncertainties in the orbital parameters are degenerate with the signal. We discuss the importance of the number of transits further below.

\subsection{Monte Carlo generation of multi-planet systems}
To obtain statistics on the expected detection frequency of secondary planets in existing circumbinary systems we conduct a Monte Carlo simulation of 10,000 possible second planets in each of the 6 detected systems. To increase our statistics we include Kepler 47 twice: once using Kepler 47b as the inner planet, and once using the probable orbital location of Kepler 47d, the so-called orphan transit \citep{Orosz:2012sp}. Thus we have effectively 7 systems. Out of 7 systems, we generously count 2 ``outer planet" detections (both in Kepler 47)\footnote{We do not include Kepler 47c as an interior planet as any planet outside of this would be undetectable in the current data}.

We specify a distribution of $\beta$'s as in \cite{Fang:2013}, using a Rayleigh distribution, with a $\sigma = 15$ for a packed populations, and $\sigma = 30$  for a sparse population. We truncate the distribution at a larger value than these authors of $\beta = 7$, because closer planets would be unstable about the binary. This makes our detection predictions somewhat  more conservative in that we remove the most easily detectable companions from the sample. We choose eccentricities and inclinations from Rayleigh distributions with $\sigma= 0.05, 3^\circ$. Note that moderate changes in these values effect the outcome very little. We assume that planetary radii follow a power law distribution in radius (scaling as $R_p^{-1.97}$), which fits well the planet population from $3R_\oplus<R_p<20 R_\oplus$ for periods between $3-50$ days \citep{Youdin:2011}. To convert radii to masses (necessary for TTV calculations), we use the fit to Earth and Saturn from  \cite{Lissauer:2011fy} where $M_p = (R_p/R_\oplus)^{2.06}M_\oplus$. For each of the seven systems, we randomly draw planets from these 
distributions, calculate the transit probability, and based on this probability count the number of expected transiting secondaries. We also calculate the magnitude of the TTV, counting all above 20 minutes as detectable.

\begin{figure*}
   \centering
     \includegraphics[scale=0.6]{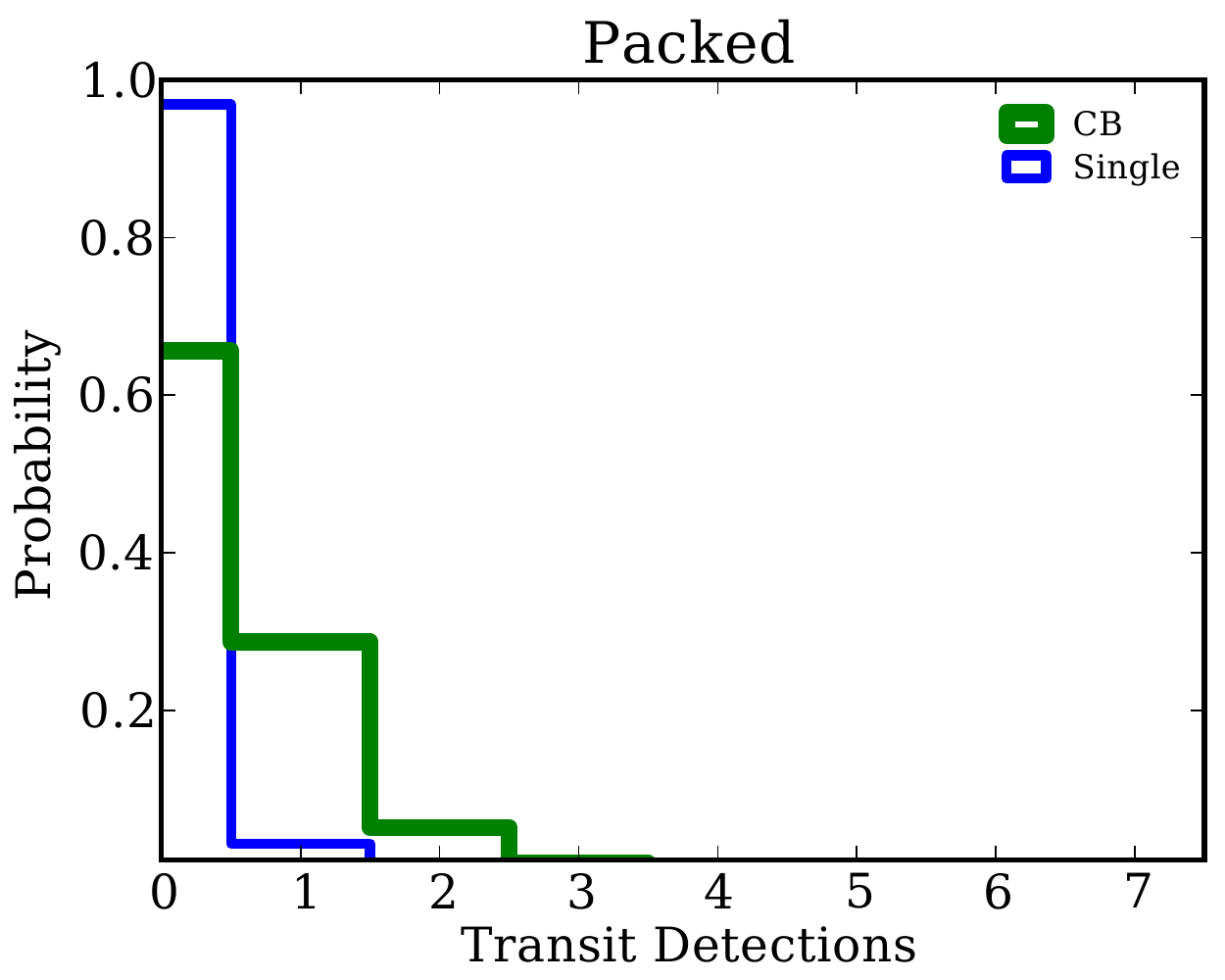} 
      \includegraphics[scale=0.6]{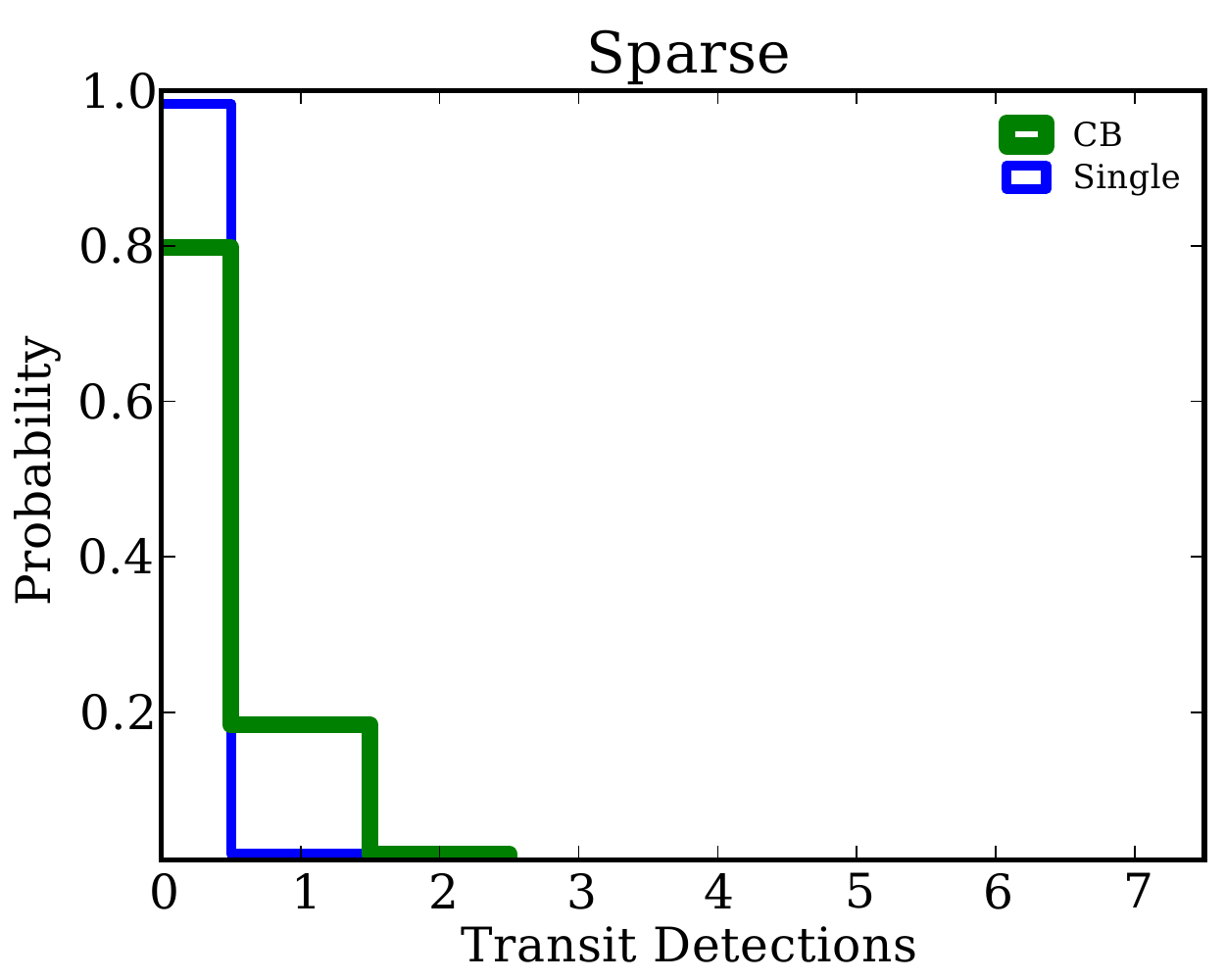} 
       \includegraphics[scale=0.6]{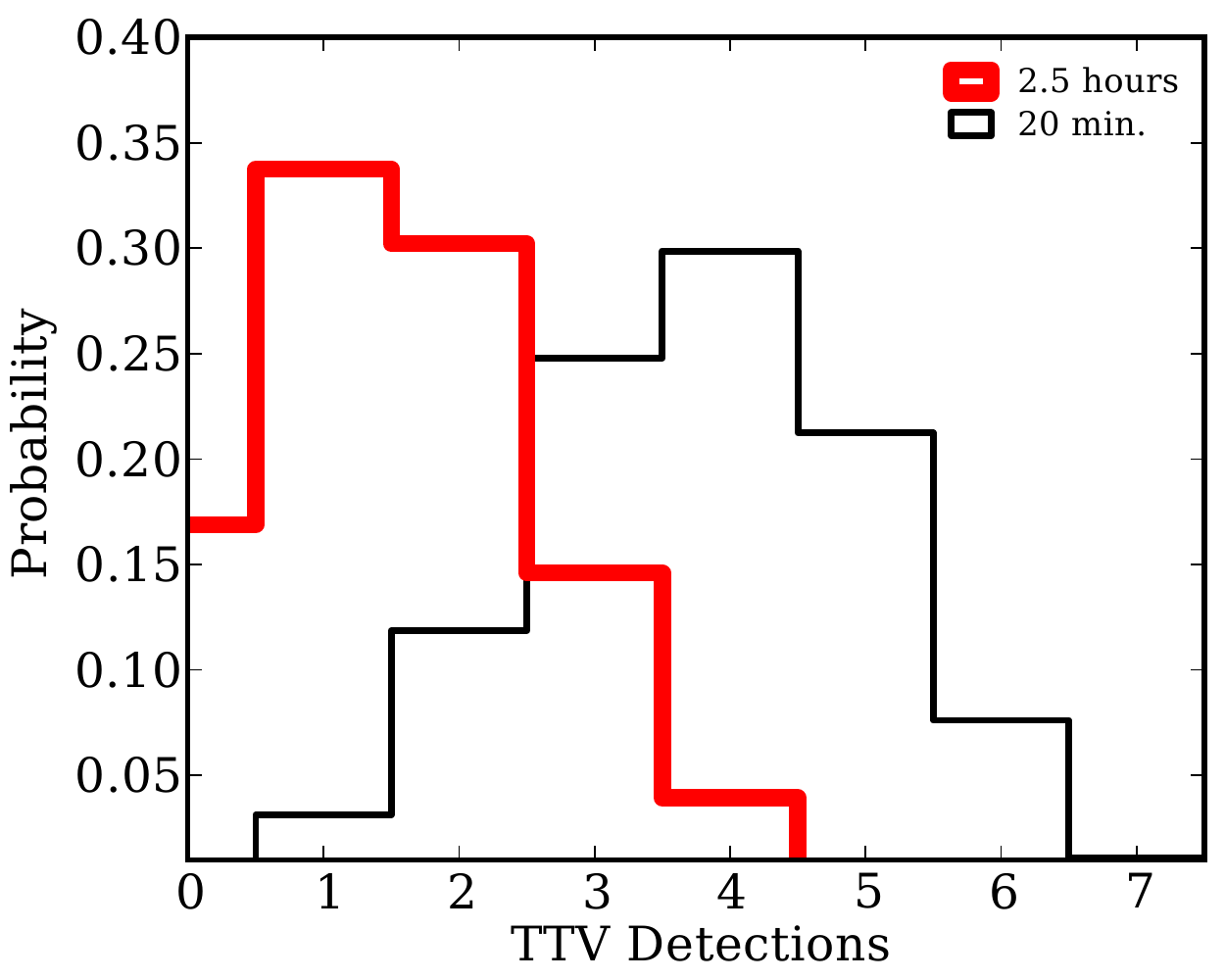}
              \includegraphics[scale=0.6]{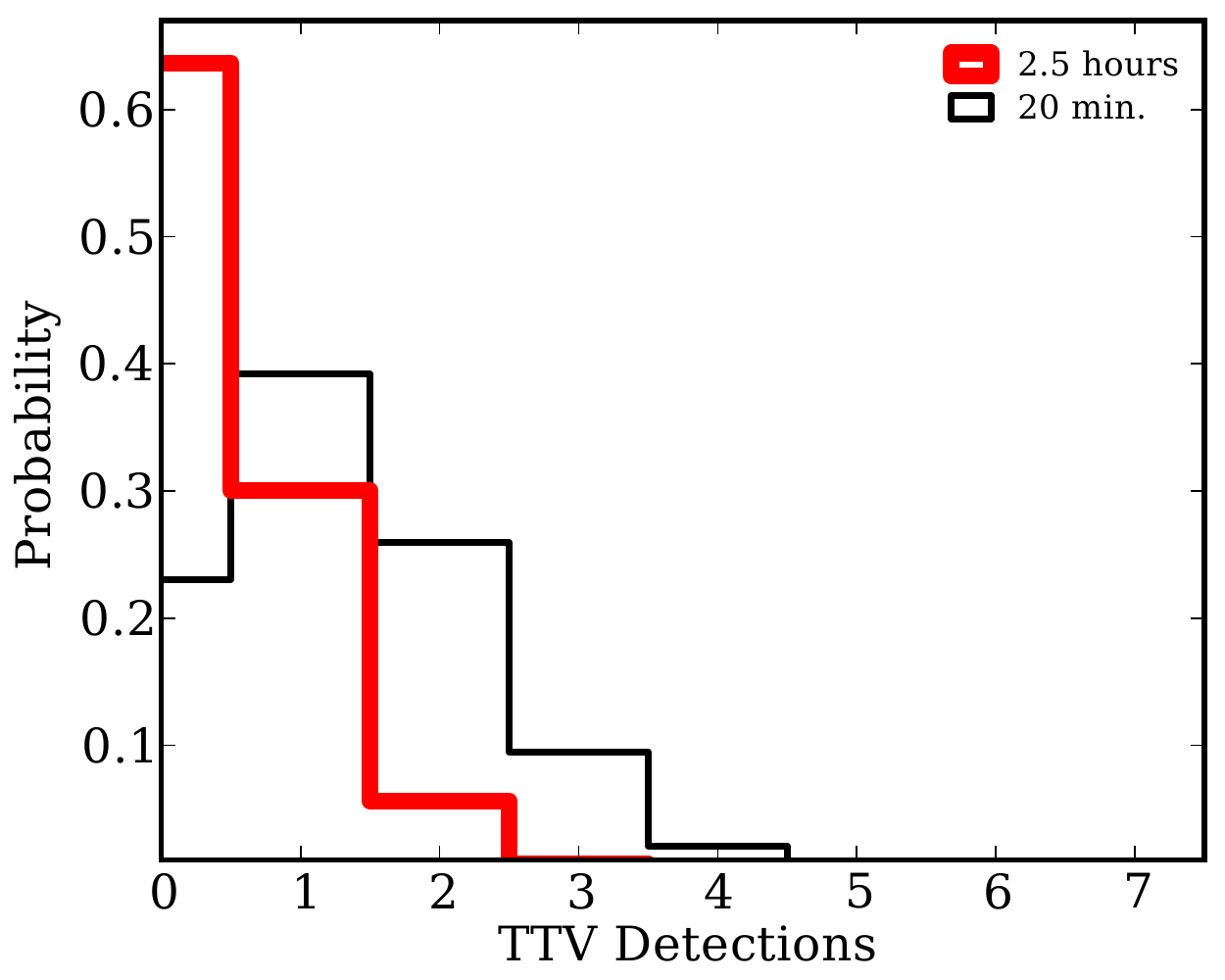}
   \caption{Probability of detecting some number of two planet systems by either transits (top) or through TTVs (bottom) for packed  (left) and sparse (right) planetary systems. We compare the transit probability for a single star and the binary  to demonstrate the enhanced detection probability for CBPs.}
\label{transit_probs}
\end{figure*}

\subsection{Results}
In \Fig{transit_probs} we show the normalized histogram distribution of the number of expected secondary detections out of 7 total systems via both transits and TTVs in both packed and sparse systems. Based on our sensitivity estimate, TTVs are the preferred detection method for secondary planets, and thus the more efficient way to differentiate between the two dynamical states. As noted above, based on our estimate, Kepler 47d should have a detectable TTV, although given the limited number of transits, characterization may be challenging.

We do not claim TTV detections will allow for the determination of a secondary planet's properties. This would require far more transits than feasible for such long period systems. It remains uncertain how many transits will be needed to  separate  degeneracies in the orbital properties of the inner planet from true TTVs. For systems like Kepler 47, where the inner planet has a period of ~50 days, 2 -3 years of data would provide 10-20 transits. Kepler 47 b was well characterized with 18 transits. Thus characterization is possible in the course of a mission timeline. Ideally one would like to observe the full oscillation period of the TTV \citep{Nesvorny:2008,Meschiari:2010}. In practice detecting even half of an oscillation might be sufficient to set a lower limit on the magnitude of the TTV.

To calculate the sample size needed to distinguish between a population of packed systems and a population of sparse systems, we use the mean detection probabilities for the packed and sparse cases ($P_p$~and $P_s$).  For transits, these rates are $P_p \sim 0.06$~and $P_s \sim 0.03$, for TTVs, they are $P_p \sim 0.55$~and $P_s \sim 0.18$.  We use Monte Carlo simulations to determine the chance that a measurement of $n$~systems will have $k$~detections.  We say that we can distinguish packed and sparse populations at $1~\sigma$~if at least $68\%$~of the Monte Carlo realizations have $k$~such that we correctly prefer one model to the other by at least $68\%:32\%$~(and similarly for $2~\sigma$~and $3~\sigma$~with $96\%$~and $99.7\%$~respectively).  Using transit detections requires 140, 1027, and 2132~systems for $1 \sigma$, $2 \sigma$, and $3 \sigma$~confidence, while using TTVs requires only 5, 34, and 75 systems for $1 \sigma$, $2 \sigma$, and $3 \sigma$~confidence. 

At present, there have been no reports of TTVs in any of the systems. Despite the $\sim 99\%$ probably of detecting at least one $>20$ minute duration TTV in the current sample if the systems are packed, we cannot yet draw conclusions about the configuration with out a more systematic investigation of the detectability of TTVs. Such an investigation will be the subject of future work. If the true TTV sensitivity is closer to 2.5 hours, the detection probabilities shrink by roughly a factor of three: In this case, the likelihood of no TTV detections if the population is packed is $\sim 16\%$, and the number of systems needed to differentiate a population of packed systems from a population of sparse systems is 4, 55, and 120 for 1, 2, and 3 $\sigma$~confidence respectively.
These calculations suggest that continued monitoring of even a few circumbinary systems in order to achieve better orbital parameters might provide substantial insight into the dynamical properties of the systems.

\section{Discussion}\label{sec:discussion}
Understanding the stable configurations of circumbinary planets, and comparing with observations, will provide invaluable insight into the formation of these complex systems. The mere existence of planets around binaries places strong constraints on the formation of the binary itself, and on the planet's natal disk. If planet formation in such disks occurs on the same timescale as around a single star (and there is no reason to think the time requirements would be less stringent), the binary must arrive at its current orbital state  in a manner that does not disrupt a circumbinary disk, or do so sufficiently early in the formation process as to permit the regrowth of a massive circumbinary disk \citep{Throop:2008}. 

Observations suggest that massive circumbinary disks are much rarer than those around single stars and wide binaries \citep{Harris:2012}. Moreover, the very closest binaries show a bi-modal distribution of disk properties: in a survey of Taurus with the SMA, most tight binaries have, if anything, disks which are too tenuous to detect. The remaining few have disks that are at the high end of the mass function of single stars. Perhaps these are the planetary system progenitors? It is interesting to note that the properties of single stars do not converge with very tight binaries, suggesting that the formation of these systems is quite different.

We have demonstrated that the stability of two-planet, circumbinary systems is distinct from that of either two or three planet single star systems, where the minimum planet separation $\beta$ is an increasing function of the inner planet separation. Depending on planet mass, the binary case asymptotes to the single star analytic two-planet results at inner-planet separations of order 1.5-2 $a_{\rm crit}$. When the inner planet is closer to $a_{\rm crit}$, the critical intra-planet spacing is of order $\beta = 5-7$.

Kepler 47 is dynamically packed if the third planet exists, but the known population of circumbinary planets is too small to robustly distinguish between a generally packed population and a generally sparse population.  An increase in the sample of circumbinary systems by roughly an order of magnitude may be sufficient to determine the typical dynamical state of such systems, but with the recent demise of the high-precision era of {\it{Kepler}}, the addition to the sample will be slow. Continued monitoring of the existing systems may prove helpful if additional transits are detectable. As noted by \cite{Borucki:1984}, eclipsing binaries provide excellent targets for the detection of transiting planets. Moreover, such planets are more likely to be found on wider, Earth-like orbits.  

We suggest that future missions such as TESS include significant numbers of eclipsing binary systems among their targets in the areas with continuous coverage. If 20 transits are sufficient to distinguish between degenerate orbital parameters and TTVs, then a 3 year mission may well be long enough to characterize the dynamical state of any discovered systems.  Moreover, TESS-like missions, which target brighter stars, are particularly valuable because these systems can be followed up from the ground.

{\it acknowledegments} We thank P. Armitage and W. Welsh for valuable discussions, and the referee for a careful review that substantially improved this paper. Support for this work was provided by NASA through Hubble Fellowship grant \#HF-51306.01 awarded by the Space Telescope Science Institute, which is operated by the Association of Universities for Research in Astronomy, Inc., for NASA, under contract NAS 5-26555. AS is supported by the European Union through ERC grant number 279973.

\end{document}